\documentclass[letterpaper]{IEEEtran}

\usepackage{cite}
\usepackage{amsmath,amssymb,amsfonts}
\usepackage{gensymb}
\usepackage{textcomp}
\usepackage{xcolor}
\usepackage{flushend}
\usepackage{algorithm,algpseudocode}
\usepackage{listings}
\DeclareMathOperator*{\argmax}{argmax} 
\usepackage{mathptmx}
\usepackage{graphicx}
\usepackage{textcomp}
\def\BibTeX{{\rm B\kern-.05em{\sc i\kern-.025em b}\kern-.08em
    T\kern-.1667em\lower.7ex\hbox{E}\kern-.125emX}}
\usepackage{makecell}

\usepackage[english]{babel}
\usepackage{amsthm}
\theoremstyle{definition}

\newtheorem{theorem}{Theorem}
\newtheorem{lemma}{Lemma}
\newtheorem{corollary}{Corollary}
\usepackage{caption}
\usepackage{subcaption}
\usepackage{mwe}

\begin{document} 

\title{
Robust Sensor Placement for Poisson Arrivals with False-Alarm–Aware Spatiotemporal Sensing
\thanks{*Corresponding author is Mingyu Kim.\\This work was supported in part by the Office of Naval Research under Grant N00014-20-1-2845.}
}

\author{Mingyu Kim$^{1*}$, Pronoy Sarker$^{1}$, Seungmo Kim$^{1}$, Daniel J. Stilwell$^2$, and Jorge Jimenez$^3$ \\ 
$^1$Department of Electrical and Computer Engineering,
Georgia Southern University, Statesboro, GA, USA\\

$^2$Bradley Department of Electrical and Computer Engineering,
Virginia Tech, Blacksburg, VA, USA\\

$^3$ Johns Hopkins University Applied Physics Laboratory, Laurel, MD, USA\\
$^*$mkim@georgiasouthern.edu}

\maketitle


\begin{abstract}
This paper studies sensor placement when detection performance varies stochastically due to environmental factors over space and time and false alarms are present, but a filter is used to attenuate the effect. We introduce a unified model that couples detection and false alarms through an availability function, which captures how false alarms reduce effective sensing and filtering responses to the disturbance. Building on this model, we give a sufficient condition under which filtering improves detection. In addition, we derive a coverage-based lower bound on the void probability. Furthermore, we prove robustness guarantees showing that performance remains stable when detection probabilities are learned from limited data. We validate the approach with numerical studies using AIS vessel-traffic data and synthetic maritime scenarios. Together, these results provide theory and practical guidance for deploying sensors in dynamic, uncertain environments.
\end{abstract}


\section{Introduction}
\label{sec:introduction}

We study optimal sensor placement for control, surveillance, and monitoring when only a limited number of sensors or mobile robots can be deployed. The goal is to choose locations that reliably detect all events of interest. Classical formulations often assume static and deterministic sensing and then optimize coverage, mutual information, or observability \cite{ChangTanXing09,Krause06,KimSPIE2023}. These assumptions are convenient but rarely hold in practice.

In marine and subsurface settings, sensing quality varies over space and time due to turbidity, sediment burial, and biofouling \cite{ChangTanXing09,Song2023TurbiditySpectral,Delgado2021Antifouling,Delauney2010Biofouling}. Additional environmental effects, such as sediment properties and seafloor interaction, further influence sensing and detection, as shown in recent seabed-embedment studies \cite{shrestha2024assessing}. In aerial applications, atmospheric and illumination factors, such as cloud cover, haze, strong sunlight, rain, and smoke, likewise modulate detection performance across space and time \cite{CampbellWynne2011,Schowengerdt2007,Colomina2014UAS}.

These effects change both detection probabilities and false-alarm rates. Filtering methods such as Kalman and particle filters can improve effective detection by exploiting temporal correlation \cite{Kalman1960,Poor1994DetectionEstimation,Gustafsson2010SensorFusion}. If overused, they can also raise false alarms \cite{Flammini10,Oduah25}. False alarms consume operator attention and vehicle time, which reduces the network’s effective availability \cite{LaymanControlled23}. In sensor placement, the interplay between detection gains and resource costs under false alarms and filtering remains poorly understood. An exception is \cite{beard2017void}, which employs void probability with false alarms for multi-object tracking. Unlike \cite{beard2017void}, our work targets \emph{sensor placement} under stochastic, spatio-temporal sensing with a guaranteed performance guidance.

Prior research provides several relevant foundations. Earlier work established Poisson-based spatial sensing models and optimal placement principles for static environments \cite{KimAccess2023,KimSPIE2023}. Follow-on studies examined stochastic target trajectories and barrier-coverage behavior \cite{KimBarrier2025}, improved approximation of spatial sensing performance in challenging seabed acoustic environments \cite{kim2025improved}, and explored data-driven characterizations of maritime uncertainty and vessel behavior \cite{jimenez2025optimal,jimenez2023optimizing}. Additional recent work investigated Poisson-based outlier detection in seabed sensing networks \cite{kim2025outlier}. These results collectively highlight the importance of modeling spatial, temporal, and environmental variability when designing real-world sensor networks.

For safety-critical operations like port security and persistent surveillance, the requirement is \emph{perfect detection}. Under Poisson arrivals, this is captured by the \emph{void probability}, the probability that no target goes undetected. Prior void-probability analyses typically assume static sensing and ignore false alarms, which limits their applicability.

We model target arrivals with a Cox process so that spatial, temporal, and environmental uncertainty are represented by a random intensity. In particular, we adopt a log--Gaussian Cox process (LGCP) where the logarithm of its intensity function is modelled as a Gaussian process \cite{Kingman1993,Moller1998,BrixDiggle2001,Diggle2013}. For data-driven studies with AIS vessel traffic, we estimate the intensity using integrated nested Laplace approximation (INLA) \cite{bachl2019inlabru,rue2017bayesian}.

To account for false alarms, we introduce an \emph{availability} function that couples detection with false-alarm behavior and filtering and quantifies the loss of effectiveness from resource drain. This function incorporates the dynamics of the external factors. We then ask when filtering helps rather than hurts and provide a verifiable condition for that regime.

Furthermore, with these stochastic target and sensing models, we search for the near-optimal sensor locations by applying greedy selection, which is computationally efficient. For performance-guarantee analysis, we derive a lower bound on void probability with a coverage condition that guarantees strictly better than the classical lower bound $1 - 1/e$ of the optimality. Finally, we establish finite-sample robustness to explore practical sensor placement considering false alarms and filtering under a dynamical environment. When detection probabilities are learned from limited data, concentration results and robust optimization ideas \cite{hoeffding1963,benTal2009,rahimian2019} ensure that placement quality degrades in a controlled way, and results from adaptive and noisy submodularity \cite{golovin2011,singla2016,hassidim2017} justify stable near-optimal performance.

\subsection*{Contributions}
We develop a framework that couples detection, filtering, and false-alarm modeling in space and time, yielding near-optimal sensor locations that maximizes void probability for uncertain targets. The contributions are: (i) a sensor model that introduces an availability function to capture the loss of effectiveness when false alarms occur while using filtering (ii) a condition that specifies when adjusting the filter setting actually improves detection, reduces the number of missed targets, and raises void probability (iii) a new lower bound on void probability that comes with a clear threshold, which strictly outperforms $1-1/e$ of the optimality depending on the operating coverage (iv) robustness guarantees showing that even with limited data, estimated detection rates concentrate around the truth, errors propagate in a controlled way, and sensor placements remain stable with a small penalty.

The rest of the paper is organized as follows. Section~\ref{sec:model} formalizes the spatio-temporal sensing setting, introduces environmental variability, filtering and false alarms via an availability model. Section~\ref{sec:theory} presents analysis and theoretical guarantees of this new objective function. In Section~\ref{sec:numerical-results}, we show simulations that illustrate the effectiveness of our framework using ship traffic data with synthetic stochastic and spatiotemporal varying environmental data. We conclude with key takeaways and possible extensions, and the Appendix provides proofs and additional technical details.

\section{Problem Formulation}\label{sec:model}

We study the problem of deploying a limited number of sensors to maximize the probability of detecting all targets in a spatio--temporal stochastic environment. Our formulation integrates three key aspects: stochastic target arrivals over space and time, sensing performance that varies with environmental conditions and filtering, and the operational cost of false alarms.

\subsection{Spatio-Temporal Stochastic Target Model}

Target arrivals are modeled by a log-Gaussian Cox process (LGCP) defined over the space--time domain $\Psi\times\mathcal{T}$. The intensity function is
\begin{align*}
\log(\lambda(s,t)) \sim GP(\mu(s,t),\Sigma(s,s',t,t'))
\end{align*}
where $s\in\Psi$ denotes location and $t\in\mathcal{T}$ denotes time.  
The expected total number of arrivals is
\begin{align*}
\Lambda = \int_{\mathcal{T}}\int_{\Psi}\lambda(s,t)\,ds\,dt
\end{align*}

\subsection{Sensor Model with Stochastic Performance and Filtering}

Unlike classical models with fixed detection probabilities, sensor performance depends on both environmental variability and filtering decisions. For a sensor $i$ placed at $a_i \in \Psi$, the probability of detecting a target at $(s,t)$ is
\begin{align*}
p(s,t,a_i;\theta,\omega) \in [0,1]
\end{align*}
where $\theta(s,t,a_i)$ is a filtering parameter (e.g., threshold or gain). For simplicity, we use $\theta_i=\theta(s,t,a_i)$. 
We model environment–induced sensing loss with a bounded stochastic field $\omega(s,t)\in[0,1]$ that modulates the effective detection probability over space and time; smaller values indicate stronger attenuation (e.g., turbidity, sediment burial, cloud cover, rain). We obtain $\omega$ by passing a latent spatio–temporal field $Z$ through a squashing function $\sigma:\mathbb{R}\to[0,1]$:
\begin{align*}
\omega(s,t) &= \sigma\!\big(Z(s,t)\big)
\end{align*}
The latent environmental field is modeled as a Gaussian process,
\begin{align*}
Z(s,t) \sim GP\big(\mu_Z(s,t),\, \Sigma_Z\big((s,t),(s',t')\big)\big),
\end{align*}
where $\mu_Z$ encodes the mean spatio–temporal trend and $\Sigma_Z$ is a covariance kernel (e.g., separable Matérn or squared–exponential) that captures spatial and temporal correlation structure.

\subsection{False-Alarm–Aware Availability}

False alarms drain operational resources and effectively reduce sensor availability.  
We capture this with an availability function
\begin{align*}
\alpha:[0,\infty)\to(0,1], \qquad \alpha(0)=1, \quad \alpha'(\chi)\le 0
\end{align*}
where $\chi(s,t;\theta_i,\omega)\ge 0$ denotes the false-alarm rate.  
The \emph{effective detection probability} for sensor $i$ is then
\begin{align*}
\tilde{p}(s,t,a_i;\theta_i,\omega)
= \alpha(\chi(s,t;\theta_i,\omega))\,p(s,t,a_i;\theta_i,\omega)
\end{align*}
For simplicity, we suppress the arguments in parentheses and use the shorthand notation 
$\tilde{p}_i=\tilde{p}(s,t,a_i;\theta_i,\omega)$,  $p_i=p(s,t,a_i;\theta_i,\omega)$, and 
$\alpha(\chi)=\alpha(\chi(s,t;\theta_i,\omega))$ for the rest of the paper.

\subsection{Undetected Targets and Void Probability}

Given a set of sensors $\mathbf{a}$ called \emph{sensor network}, the probability that a target at $(s,t)$ is undetected is
\begin{align*}
\prod_{i\in \mathbf{a}}\big(1-\tilde p_i\big)
\end{align*}
The expected number of undetected targets by the sensor network $\mathbf{a}$ is
\begin{align*}
U(\mathbf{a}) = \int_{\Psi\times\mathcal{T}} \lambda(s,t)\,
\prod_{i\in \mathbf{a}}\big(1-\tilde p_i\big)\,ds\,dt
\end{align*}
The probability that no target is missed (void probability) is
\begin{align*}
\nu(\mathbf{a}) = \mathbb{E}_{\lambda,\omega}\!\left[\exp\!\big(-U(\mathbf{a})\big)\right]
\end{align*}

\subsection{Near-Optimal Sensor Placement}

The optimal placement problem is
\begin{align*}
\mathbf{a}^{\star}=\argmax_{\mathbf{a}\subseteq\mathcal{L},\;|\mathbf{a}|\le K}\;\; \nu(\mathbf{a})
\end{align*}
where $\mathcal{L}$ is all possible sensor locations and $K$ is the number of sensors.  
Directly optimizing $\nu(\mathbf{a})$ is intractable due to stochasticity, so we approximate it using the lower bound from Jesens's inequality.
By Jensen’s inequality,
\begin{align*}
\nu(\mathbf{a}) \;\ge\; \exp\!\big(-\bar U(\mathbf{a})\big) 
\end{align*}
where 
\begin{align*}
    \bar U(\mathbf{a}) = \mathbb{E}_{\lambda,\omega}[U(\mathbf{a})]
\end{align*}
Using this lower bound as the alternative objective function, we can reformulate the optimization problem
\begin{align*}
\hat{\mathbf{a}}^{\star}
= \argmax_{\mathbf{a}\subseteq\mathcal{L},\;|\mathbf{a}|\le K}\;\; \exp\!\big(-\bar U(\mathbf{a})\big)
\end{align*}

\section{Analysis and Guarantees under Stochastic Sensing and False Alarms} \label{sec:theory}

In this section, we analyze structural properties of the sensing model, the role of filtering under false alarms, and the robustness of greedy placement when only finite data are available. The key results are: (i) filtering is provably beneficial under a simple condition, (ii) when the coverage with greedy selection method satisfies a simple condition, it strictly outperforms the $1-1/e$ of the optimality and (iii) greedy placement remains near-optimal even with estimation noise. Finally, we establish a lower bound on the void probability that connects algorithmic guarantees. Proofs of all results in this section are provided in the Appendix.

\subsection{Filtering Benefit Considering False Alarms}
While filtering can improve detection, imperfect filters may elevate false alarms. We therefore seek a condition under which filtering yields a net performance benefit.

\begin{theorem}[Sufficient Local Condition]
Assume $p_i>0$ and $\alpha(\chi)>0$. For a sensor $i\in\{1,\dots,m\}$, if $\tilde p_i$ is differentiable in $\theta_i$, then
\[
\frac{\partial_{\theta_i} \tilde p_i}{\tilde p_i}
=\frac{\partial_{\theta_i} p_i}{p_i}
+\frac{\alpha'(\chi)}{\alpha(\chi)}\,\partial_{\theta_i} \chi
\]
If, pointwise on $(s,t)$,
\begin{align}
\frac{\partial_{\theta_i} p_i}{p_i}
\;\ge\;-\frac{\alpha'(\chi)}{\alpha(\chi)}\,\partial_{\theta_i} \chi \label{eq: sufficient}    
\end{align}

then increasing $\theta_i$ improves effective detection $\tilde p_i$, decreases the expected number of undetected targets $\bar U(\mathbf{a})$, and increases the void probability $\nu(\mathbf{a})$.\label{the: sufficient cond}
\end{theorem}

\noindent This implies that filtering should be adjusted only when the local gain in raw detection exceeds the penalty induced by false-alarm sensitivity. This condition provides a rigorous guideline for adaptive sensor settings in dynamic environments. The proof is in the Appendix.

\subsection{Coverage–Oriented Void Probability}
The preceding result established when filtering provides a net performance gain by balancing detection and false alarms in dynamic environments. However, due to environmental variability and imperfect sensing, applications where missed detections are unacceptable often require short sensing intervals to ensure sufficient spatial coverage. As additional sensors are deployed, the expected number of undetected targets cannot increase; yet the marginal gain diminishes because of overlapping sensing regions. This diminishing-returns behavior is the hallmark of submodularity. We define the coverage of a sensor set $\mathbf{a}$ as
\[
C(\mathbf{a}) \triangleq \Lambda(\Psi,\mathcal{T}) - \bar U(\mathbf{a})
\]
the expected number of detected targets. The set function $C$ is monotone submodular \cite{KimAccess2023}. By Nemhauser \emph{et al.} \cite{nemhauser1978analysis}, greedy placement under a cardinality constraint satisfies
\begin{equation}
C(\mathbf{a}_{\text{greedy}}) \;\ge\; (1-1/e)\, C(\mathbf{a}^\star)
\label{eq:greedy-coverage}
\end{equation}
where $\mathbf{a}_{\text{greedy}}$ and $\mathbf{a}^\star$ denote the greedy and optimal placements, respectively. Equivalently,
\begin{equation}
\bar U(\mathbf{a}_{\text{greedy}})
\;\le\; \tfrac{1}{e}\,\Lambda(\Psi,\mathcal{T})
+\big(1-\tfrac{1}{e}\big)\,\bar U(\mathbf{a}^\star)
\label{eq:greedy-U}
\end{equation}

\noindent From \eqref{eq:greedy-U}, the void probability under greedy placement admits the lower bound
\begin{align}
\nu(\mathbf{a}_{\mathrm{greedy}})
\;\ge\;
\exp\!\Big(
-\tfrac{1}{e}\,\Lambda(\Psi,\mathcal{T})
-\big(1-\tfrac{1}{e}\big)\,\bar U(\mathbf{a}^\star)
\Big)
\label{eq:lower-bound-coverage-void}
\end{align}
For comparison, the approximate objective in \cite{KimAccess2023} yields
\begin{align}
\nu(\mathbf{a}_{\mathrm{greedy}})
\;\ge\;
(1-1/e)\,\exp\!\big(-\bar U(\mathbf{a}^\star)\big)
\label{eq:approx-obj-access}
\end{align}

\begin{lemma}[Dominance Threshold of Coverage-Based Bound]\label{lem:dominance-threshold}
The bound in \eqref{eq:lower-bound-coverage-void} dominates \eqref{eq:approx-obj-access} if and only if
\begin{align}
C(\mathbf{a}^\star) \;<\; -e\,\ln\!\Big(1-\tfrac{1}{e}\Big)\;\approx\;1.2468 \label{eq: dominance ineq}  
\end{align}

\end{lemma}

\noindent The right-hand side of \eqref{eq: dominance ineq} follows by taking the ratio of the lower bounds in \eqref{eq:approx-obj-access} and \eqref{eq:lower-bound-coverage-void}. The lemma thereby identifies exactly the optimal coverage in which the coverage-based guarantee is tighter.

\begin{corollary}[Strictly greater than $(1-1/e)\nu(\mathbf{a^{\star}})$]\label{cor:switching}
Under greedy placement,  if
\begin{align}
C(\mathbf{a}_{\mathrm{greedy}})\;< \;-(e-1)\,\ln\!\Big(1-\tfrac{1}{e}\Big) \label{eq: switching limit}    
\end{align}
then, 
\begin{align*}
\nu(\mathbf{a}_{\mathrm{greedy}}) > (1-1/e)\nu(\mathbf{a^{\star}})
\end{align*}
\end{corollary}

\noindent Since we cannot find the optimal coverage $C(\mathbf{a}^{\star})$, this corollary shows with repect to the greedy sensor placement. Condition \eqref{eq: switching limit} identifies a low-coverage regime in which the coverage-based certificate is strictly stronger than the classic \((1-1/e)\) of the optimality guarantee. In particular, when the greedy coverage \(C(\mathbf{a}_{\mathrm{greedy}})\) is below the threshold \(-(e-1)\ln(1-1/e)\approx 0.788\), the bound in Corollary~\ref{cor:switching} certifies
\(\nu(\mathbf{a}_{\mathrm{greedy}})>(1-1/e)\,\nu(\mathbf{a}^\star)\).
Outside this regime (higher coverage), the alternative bound may be tighter, so one should report whichever lower bound is stronger for the operating point.

\subsection{Finite–Sample Robustness}
The preceding sections established performance guarantees under exact detection probabilities. In practice, however, detection rates must be estimated from finite data. We therefore analyze how robust these guarantees remain under estimation error. Proofs are deferred to the Appendix section.

\begin{theorem}[Uniform Concentration]
For sensor $i\in\{1,\dots,m\}$ at a discretized cell $g\in\{1,\dots,L\}$, let $\widehat p_{i,g}$ denote the empirical estimate of $\tilde p_{i,g}$ from $N$ Monte Carlo samples. 
For any confidence parameter $\delta \in (0,1)$, with probability at least $1-\delta$, Hoeffding’s inequality \cite{hoeffding1963probability} implies
\[
\max_{i,g}\,|\widehat{p}_{i,g}-\tilde p_{i,g}|
\;\le\;
\varepsilon_N, \qquad
\varepsilon_N=\sqrt{\tfrac{1}{2N}\ln\tfrac{2mL}{\delta}} 
\]
where $\varepsilon_N$ is the uniform concentration error that decreases as the sample size $N$ grows, and $\delta$ controls the confidence level of the bound.
\label{the:Uniform}
\end{theorem}

\noindent This result shows that empirical estimates $\widehat{p}_{i,g}$ are uniformly close to the true detection probabilities $\tilde{p}_{i,g}$ across all sensors and cells. The error decays as the sample size $N$ increases and depends only logarithmically on the number of sensors $m$ and cells $L$. 

\begin{lemma}[Error in Undetected Targets]
For any placement $\mathbf{a}$ with at most $K$ sensors,
\[
\big|\,\widehat{U}(\mathbf{a})-\bar U(\mathbf{a})\,\big|
\;\le\; C_K\,\varepsilon_N,
\qquad
C_K \triangleq K\sum_{g} \lambda_g\,|g| 
\]
\label{lemma: error}
\end{lemma}

\noindent Here $\widehat{U}(\mathbf{a})$ denotes the empirical estimate of the undetected-target metric obtained from $N$ Monte Carlo samples.The constant $C_K$ captures the worst-case amplification of the per-cell probability error $\varepsilon_N$ from Theorem~\ref{the:Uniform}, scaling linearly with both the number of sensors $K$ and the expected total number of arrivals $\Lambda=\sum_g \lambda_g |g|$. This lemma quantifies how estimation error propagates into the mission objective, bounding the worst-case deviation in the expected number of missed targets.

\begin{theorem}[Greedy Stability with Estimates]
Let $\widehat{\mathbf{a}}_{\mathrm{greedy}}$ be the greedy placement computed from $\widehat p$. Then, with probability at least $1-\delta$,
\begin{align}
\nu(\widehat{\mathbf{a}}_{\mathrm{greedy}})
\;\ge\;
\exp\!\Big(-\frac{1}{e}C(\mathbf{a}^\star)- \bar U(\mathbf{a}^\star)
- C'_K\,\varepsilon_N
\Big) \label{eq: lower bound of finite samples}    
\end{align}

where $C'_K=O\!\Big(K\sum_g \lambda_g|g|\Big)$ \label{theo:greedy stability}
\end{theorem}

\noindent
In here, $\nu(\widehat{\mathbf{a}}_{\mathrm{greedy}})$ denotes the \emph{void probability} when sensors are placed according to the greedy solution computed from the empirical detection probabilities $\widehat{p}$. 
This theorem certifies that even when detection rates are estimated from finite samples, the greedy algorithm continues to provide strong guarantees. The additional penalty term is explicit and diminishes as $N$ grows, ensuring stability of performance under sampling uncertainty.

\begin{figure}[t]
    \centering
    \includegraphics[width=0.9\linewidth]{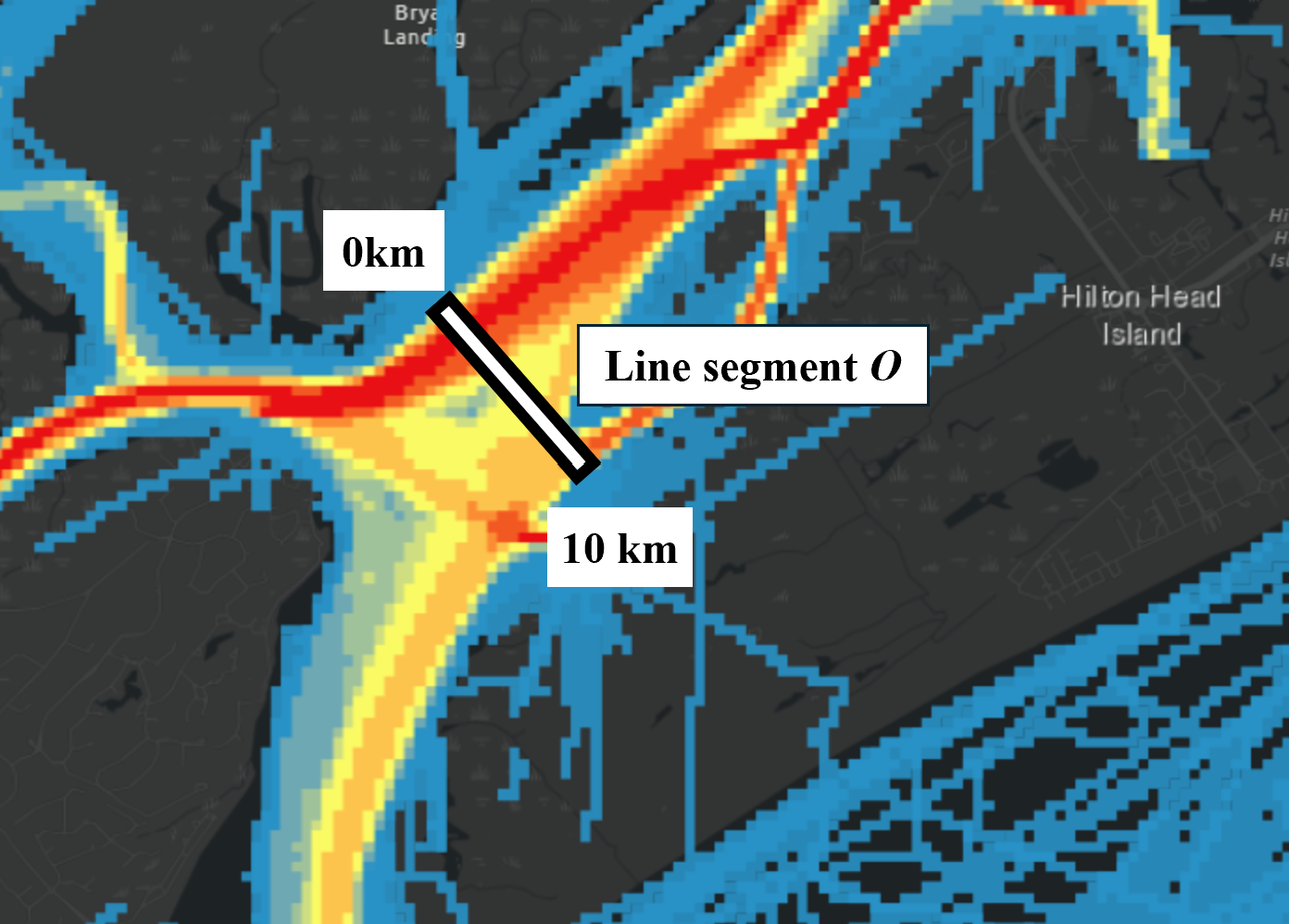}
    \caption{Ship traffic heatmap near Hilton Head Island, Georgia, USA \cite{MarineCadastreAIS}. 
    The dashed line denotes line segment $O$, used as the area of interest.}
    \label{fig:0}
\end{figure}

\section{Numerical Results}\label{sec:numerical-results}

We evaluate the framework using synthetic simulations grounded in AIS ship traffic along a representative line segment near Hilton Head Island, Georgia, USA as shown in Fig.~\ref{fig:0}. The environmental dynamics are synthetically generated, which can be replaced by data collected by the environment in real-time. The AIS data provide arrival time, location, and vessel identifiers. We work in one dimension for clarity, and the framework extends directly to two dimensions \cite{KimBarrier2025} or higher. The experiments demonstrate the efficacy of each of the theoretical results in Section~\ref{sec:theory}: (i) coverage-based void probability bound serve as practical certificate, (ii) filtering improves detection when the sufficient condition holds, and (iii) greedy guarantees remain stable under finite-sample estimation.
\begin{figure}[t!]
    \centering
    \includegraphics[width=1\linewidth]{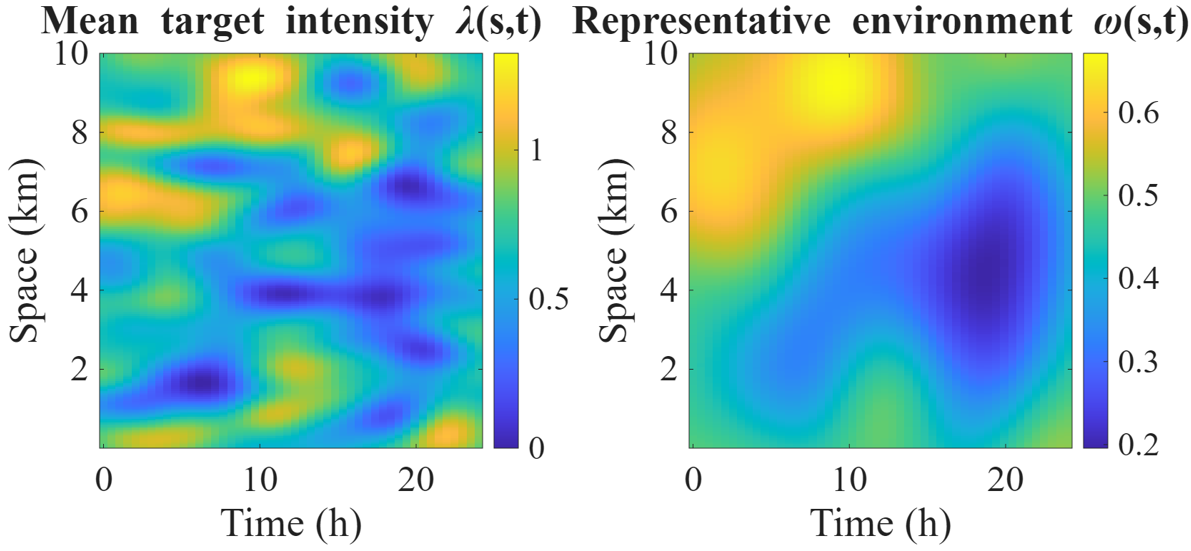}
    \caption{(Left) Posterior mean of target intensity $\lambda(s,t)$. (Right) Representative realization of the environment factor $\omega(s,t)$.}
    \label{fig:lambda_omega}
\end{figure}

\begin{figure}[t]
    \centering
    \includegraphics[width=0.95\linewidth]{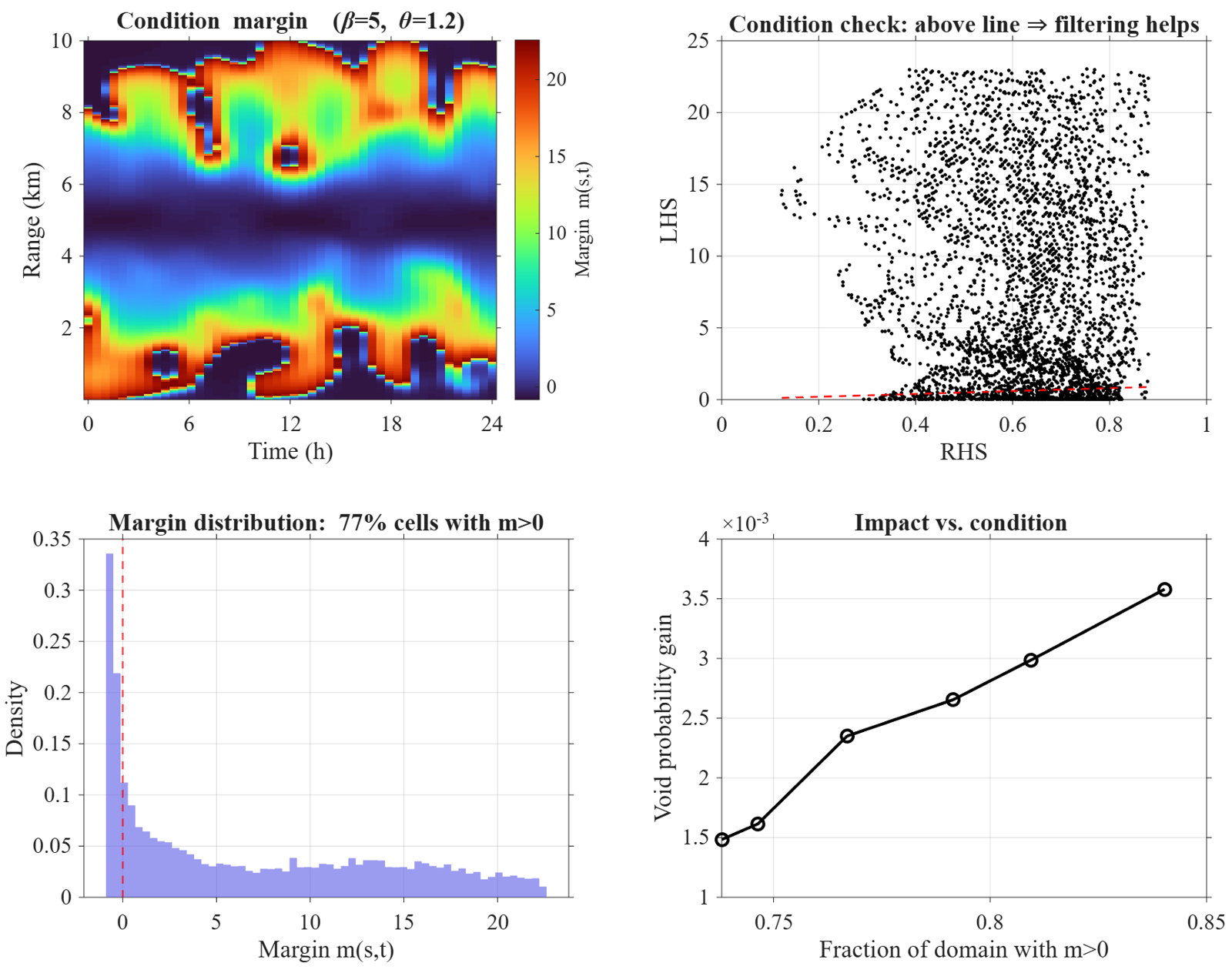}
    \caption{Filtering diagnostics: (A) margin $m(s,t)$, (B) scatter validation, (C) margin distribution, and (D) mission–level gains versus fraction of cells with $m>0$.}
    \label{fig:condition_diagnostics}
\end{figure}

We consider AIS ship–arrival data collected from January to March 2024 along a one–dimensional maritime corridor defined by line segment $O$, extending from $\big(-80.835^\circ,\;32.145^\circ\big)$ to $\big(-80.811^\circ,\;32.148^\circ\big)$ (longitude, latitude). The arc length of the segment is parameterized as $s\in\Psi=[0,10]~\mathrm{km}$ over a time horizon $\mathcal{T}=[0,24]~\mathrm{h}$. The domain is discretized into $200$ spatial bins of $50~\mathrm{m}$ and $48$ temporal bins of $30~\mathrm{min}$.

The target arrival process is modeled as a log–Gaussian Cox process (LGCP) fitted using the integrated nested Laplace approximation (INLA) \cite{bachl2019inlabru,lindgren2015bayesian}. The posterior mean intensity $\lambda(s,t)$ serves as the baseline input for numerical experiments, while $M=30$ posterior samples are employed to evaluate robustness. Fig.~\ref{fig:lambda_omega} depicts the posterior mean intensity field (left) and a representative realization of the environmental degradation factor $\omega(s,t)$ (right).

\subsection{Sensing and False–Alarm Models}
\label{sec:sensing-false-alarm}
\paragraph*{Environment-to-sensing linkage}

The environmental dynamics are modeled as a spatio–temporal Gaussian process (GP) composed with a monotone squashing function $\sigma(\cdot)$, a standard construction used in GP classification and bounded-response modeling \cite{RasmussenWilliams2006,NickischRasmussen2008}. For simulations we set $\mu_Z\equiv 0$ and use a separable squared–exponential kernel,
\begin{align}
\Sigma_Z\!\big((s,t),(s',t')\big)
= \sigma_Z^2 \exp\!\left(-\frac{(s-s')^2}{2\ell_s^2}-\frac{(t-t')^2}{2\ell_t^2}\right),
\end{align}
which is a common choice for spatio–temporal GP modeling \cite{CressieWikle2011,Banerjee2014}. We take $\sigma_Z=0.8$, $\ell_s=0.5~\mathrm{km}$, and $\ell_t=1~\mathrm{h}$. The squashing map is $\sigma(Z)=1-e^{-\beta_\omega Z}$ with $\beta_\omega=1.5$, yielding $\omega\in(0,1)$ over $\Psi=[0,10]~\mathrm{km}$ and $\mathcal{T}=[0,24]~\mathrm{h}$.

To couple $\omega$ to sensing, we then set the effective length (range) as
\begin{align}
\ell(\theta_i,\omega,s,t)
\;=\; \theta_i(s,t)\,e^{-\omega(s,t)}
\;
\label{eq:length-ell}
\end{align}
so that stronger degradation (larger $\omega$) shortens range.

\paragraph*{Detection model}
For a sensor $i$ located at $a_i$, the probability of detection is
\begin{align}
p_i(s,t,\theta_i,\omega) 
= \exp\!\left(-\frac{(s-a_i)^2}{\ell(\theta_i,\omega,s,t)}\right)
\label{eq:detect}
\end{align}
with $\ell(\cdot)$ as in \eqref{eq:length-ell}. Equation \eqref{eq:detect} recovers longer range in 
benign conditions (small $\omega$) and attenuated range under adverse conditions (large $\omega$).

\begin{figure}[t]
    \centering
\includegraphics[width=0.95\linewidth]{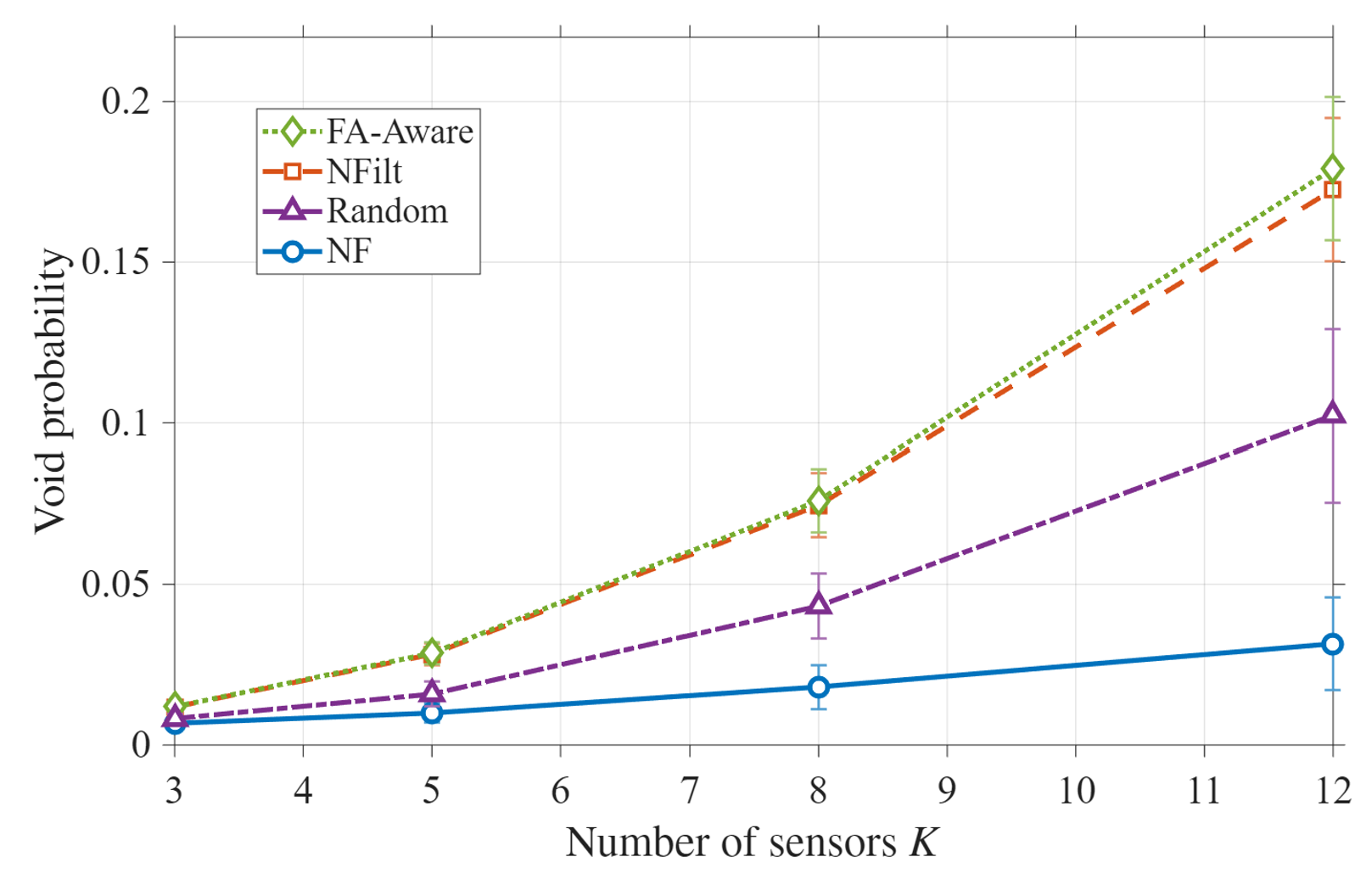}
    \caption{Void probability with number of sensors $K$ for NF, NFilt, Random, and FA–Aware with error bars $\pm 1$ std across 30 realizations.}
\label{fig:void_probability}
\end{figure}
\paragraph*{Availability and false–alarm surrogate}
The fraction of time a sensor remains operational after accounting for false alarms is
\begin{align}
\alpha(\chi)=\frac{1}{1+\beta\,\chi},\qquad \beta>0
\label{eq:availability}
\end{align}
where $\beta$ encodes the operational penalty. The false–alarm surrogate couples environment and
filter sensitivity,
\begin{align}
\chi\big(\theta_i,s,t,\omega\big)=\omega(s,t)\;\zeta\!\big(\theta_i(s,t)\big)
\label{eq:false-alarm}
\end{align}
with
\begin{align}
\zeta\!\big(\theta_i(s,t)\big)=\big(\theta_i(s,t)-1\big)^2+\xi
\end{align}
where we set the maximum filtering performance for the false alarm $\xi=0.2 \in [0,1]$ in this example.
\label{eq:zeta}
In the numerical study we use $\beta=5$. Thus, harsher conditions (larger $\omega$) increase the
false–alarm surrogate \eqref{eq:false-alarm}, reducing availability via \eqref{eq:availability}.

\subsection{Filtering Benefit Under False Alarms}
As summarized in Fig.~\ref{fig:condition_diagnostics}, panels (A)–(C) assess when filtering is beneficial via the local margin
$m(s,t)=\text{LHS}-\text{RHS}$, where $\text{LHS}=(\partial_\theta p)/p$ and
$\text{RHS}=\beta\,(\partial_\theta \nu)/(1+\beta\nu)$ in \eqref{eq: sufficient}. Panel (A) plots a heatmap of $m$ over the
$[0,10]~\mathrm{km}\times[0,24]~\mathrm{h}$ grid for $\theta=1.2$: warm regions ($m>0$) indicate filtering helps, while cool regions ($m<0$) indicate harm. Panel (B) scatters $\text{LHS}$ versus $\text{RHS}$ for $4,000$ space–time cells sampled uniformly from the grid; points above the $45^\circ$ line (red dashed) satisfy \eqref{eq: sufficient} and thus have $m\ge 0$. Panel (C) shows the distribution of $m$ and the fraction of cells with $m>0$ (vertical red dashed line).

Panel (D) connects this local diagnostic to mission performance. As $\beta$ is swept, the x–axis records the fraction of cells with $m>0$, and the y–axis shows the gain in void probability achieved when placement uses the false–alarm–aware surrogate $\alpha\,p$ instead of $p$ alone. A larger share of $m>0$ cells correlates with greater improvement.

\begin{figure}[t]
    \centering
    \includegraphics[width=0.95\linewidth]{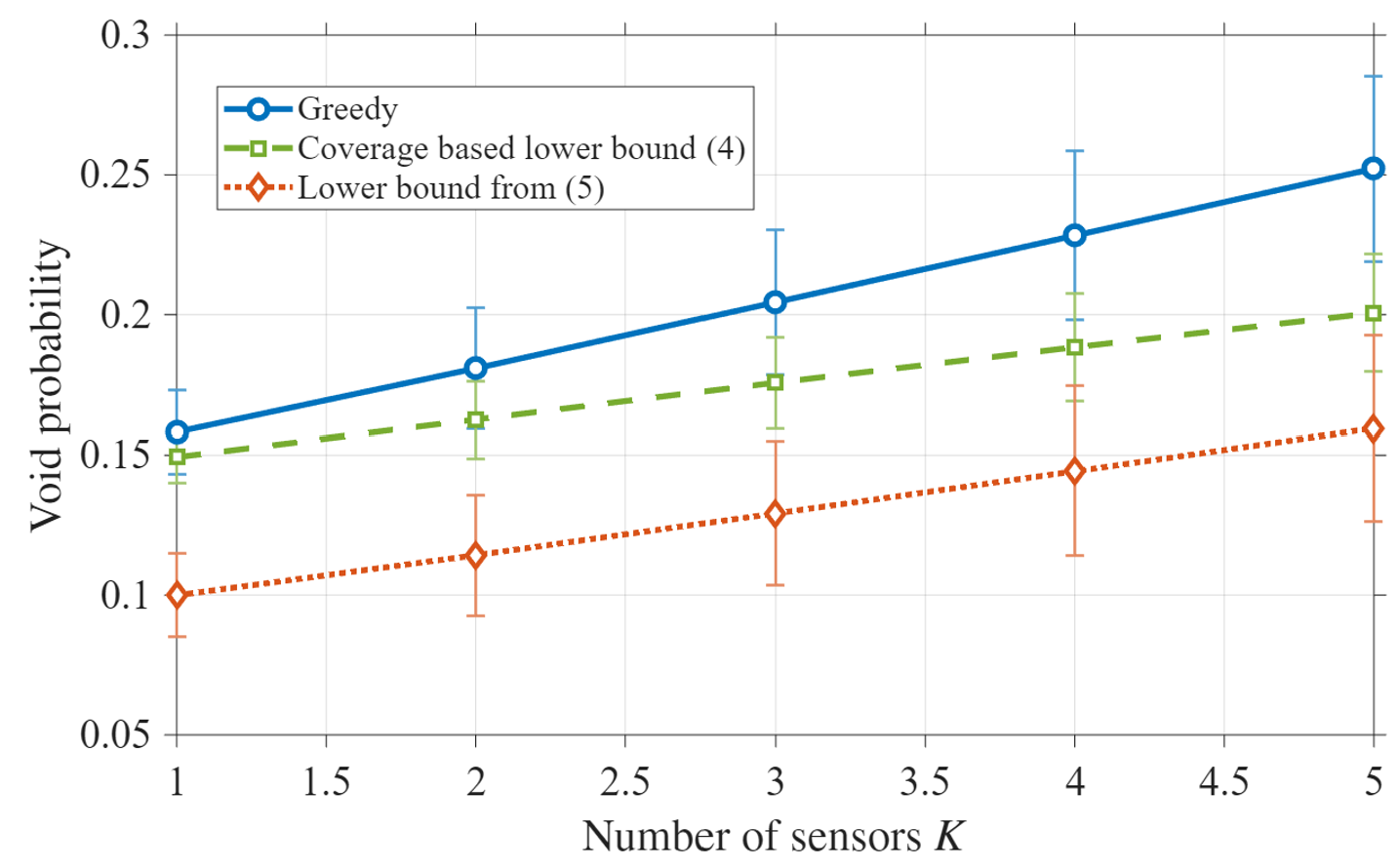}
    \caption{Void Probability varying the number of sensors: Greedy, and lower bounds from \eqref{eq:lower-bound-coverage-void} and \eqref{eq:approx-obj-access}.}
    \label{fig:bound_vs_realized}
\end{figure}

\begin{figure*}[t]
    \centering
\includegraphics[width=0.7\linewidth]{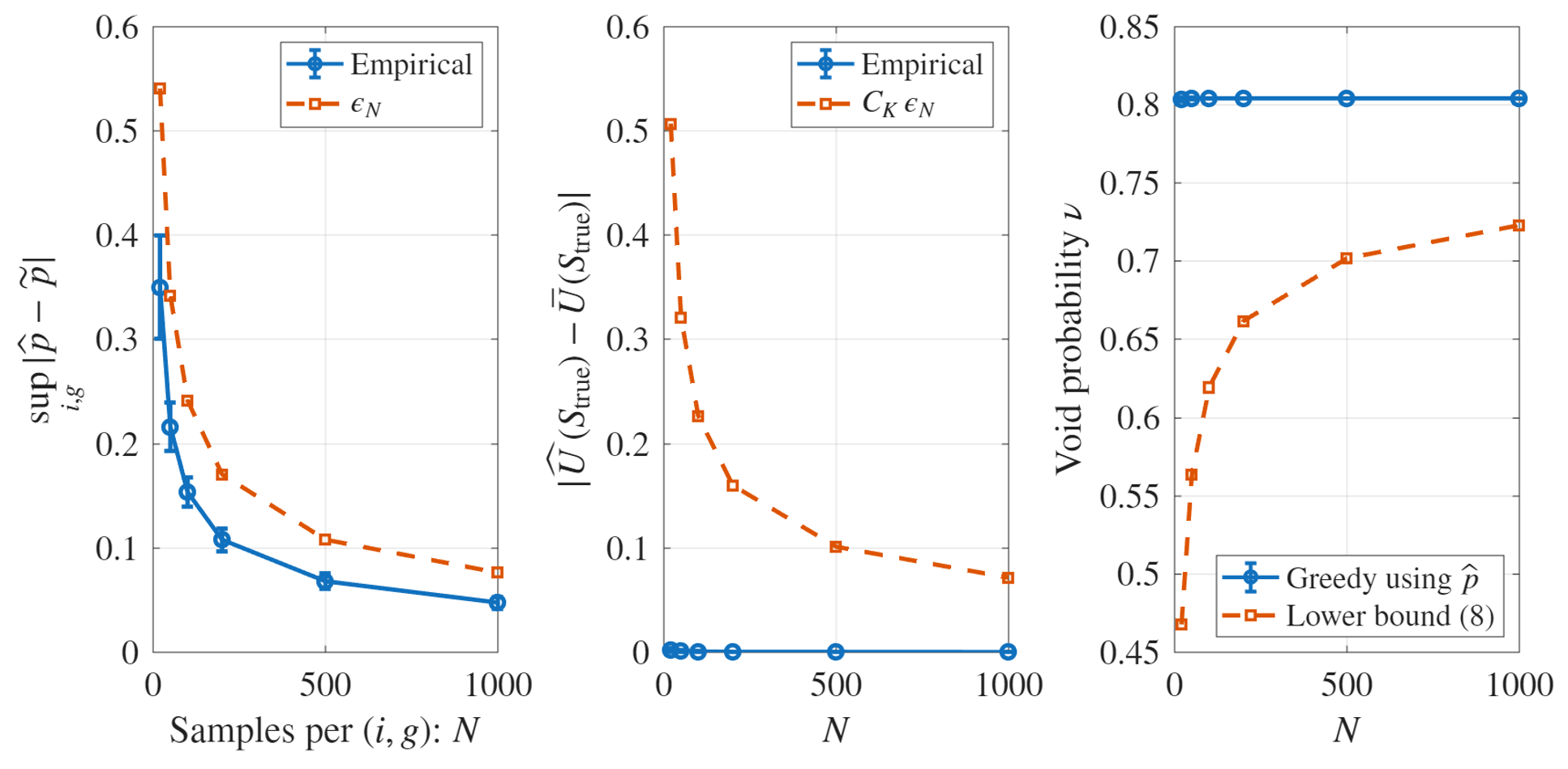}
    \caption{Finite--Sample Robustness (15-min window): Concentration, Propagation, and Greedy Stability. 
    \textbf{(left)} Empirical uniform error $\max_{i,g}\lvert\widehat p-\tilde p\rvert$ versus the Hoeffding bound $\epsilon_N$; both decay at the canonical $N^{-1/2}$ rate. 
    \textbf{(center)} Deviation in expected misses $\lvert\widehat U-\bar U\rvert$ compared against $C_K\epsilon_N$, illustrating linear propagation of estimation error to the mission metric. 
    \textbf{(right)} True void probability of the greedy placement built from $\widehat p$ compared to its finite–sample lower bound; increasing $N$ tightens the bound and improves performance.}
    \label{fig:finite_sample}
\end{figure*}
\subsection{Coverage–Oriented Void Probability}
 We compare four placement policies, chosen to isolate the impact of filtering and false–alarm awareness:
\begin{itemize}
    \item \emph{NF (no filtering awareness):} sensor location selection uses only the bare probability of detection \(p(s,t)\) and ignores availability penalties. This reflects planning that assumes ideal sensing range and uptime.
    \item \emph{NFilt (filtered, but no false alarm (FA) penalty):} selection uses the filtered range model with fixed \(\theta\) but still ignores availability; this captures designs that tune range/quality yet do not account for false–alarm–induced downtime.
    \item \emph{FA–Aware (availability–aware filtering):} selection uses the surrogate \(\alpha p\), explicitly trading off detection range with reductions in availability from false alarms; this is the proposed, operationally realistic policy.
    \item \emph{Random:} sensor locations are drawn uniformly at random over \(\Psi\); this provides a naïve baseline without modeling.
\end{itemize}
Across \(K\), FA–Aware consistently dominates NFilt, NF, and Random in realized void probability, demonstrating that incorporating availability into the planning objective yields improvements.

Fig.~\ref{fig:bound_vs_realized} compares the FA-aware curve with two computable lower bounds on the void probability. As shown, when \eqref{eq: switching limit} holds, the lower bound of \eqref{eq:lower-bound-coverage-void} is tighter.

The coverage-based bound \eqref{eq:lower-bound-coverage-void} (evaluated using \(C(\mathbf{a}_{\mathrm{greedy}})\)) and the approximate-objective bound \eqref{eq:approx-obj-access}.
Both bounds are conservative yet track the realized curve.
We define \(\tau \triangleq -e\ln\!\big(1-\tfrac{1}{e}\big)\).
By Lemma~\ref{lem:dominance-threshold}, the relative tightness aligns with the coverage level: when \(C(\mathbf{a}^\star)< \tau\) the coverage-based bound \eqref{eq:lower-bound-coverage-void} is tighter, otherwise \eqref{eq:approx-obj-access} is tighter.
Corollary~\ref{cor:switching} further provides a practical proxy test using the observed greedy coverage with a condition $\tau'\triangleq-(e-1)\,\ln\!\Big(1-\tfrac{1}{e}\Big)$: if
\(C(\mathbf{a}_{\mathrm{greedy}}) < \tau'\) \eqref{eq: switching limit},
then \(\nu(\mathbf{a}_{\mathrm{greedy}}) > (1-1/e)\,\nu(\mathbf{a}^\star)\) and the coverage-based bound is preferred. 

\subsection{Finite–Sample Robustness}
Fig.~\ref{fig:finite_sample} summarizes robustness when the true detection matrix $\tilde p$ is replaced by empirical estimates $\widehat p$ formed from $N$ samples per sensor–cell over a short (15\,min) window. \textbf{(left)} The uniform error $\max_{i,g}\lvert \widehat p_{i,g}-\tilde p_{i,g}\rvert$ closely follows (and typically lies below) the Hoeffding bound $\varepsilon_N=\sqrt{\tfrac{1}{2N}\ln\tfrac{2mL}{\delta}}$ from Theorem~\ref{the:Uniform}, confirming the canonical $N^{-1/2}$ decay. \textbf{(center)} Holding the oracle greedy set fixed, the deviation in expected misses $\lvert \widehat U(\mathbf a)-\bar U(\mathbf a)\rvert$ scales linearly with $\varepsilon_N$ and is well tracked by $C_K\varepsilon_N$ (Lemma~\ref{lemma: error}), showing that small probability errors translate into proportionally small mission error. \textbf{(right)} The realized void probability of the greedy placement built from $\widehat p$ remains above the finite–sample lower bound $\exp\!\big(-\tfrac{1}{e}\Lambda-(1-\tfrac{1}{e})\bar U(\mathbf a^\star)-C'_K\varepsilon_N\big)$ (Theorem~\ref{theo:greedy stability}); as $N$ grows, the penalty term shrinks, the bound tightens, and performance approaches the oracle. Collectively, the panels show that modest sample sizes suffice for near–oracle planning while preserving certified guarantees.

\section{Conclusion}
We introduce a framework for sensor placement that explicitly accounts for stochastic, spatio–temporal sensing variability and the operational impact of false alarms considering filtering at the same time. By modeling detection, filtering, false alarms jointly through an availability function, we captured the tradeoff between improved detection and reduced sensor availability. We establish a sufficient condition for when filtering is beneficial, derive a coverage-oriented lower bound when applying greedy selection on void probability where under a simple condition, the lower bound strictly outperforms the classical lower bound $1-1/e$ of the optimal. Furthermore, we analyze robustness under finite data, showing that performance guarantees persist even with estimation noise. Numerical studies using AIS ship-traffic data and synthetic environments validate these theoretical insights, demonstrating that false-alarm–aware filtering leads to higher void probability and more reliable performance in dynamic conditions. Future work will extend the framework to higher-dimensional domains, decentralized decision-making across sensor networks, and integration with real-time adaptive control for autonomous systems.
\appendices
\section*{Proofs of Results in Section~\ref{sec:theory}} \label{app:proofs}

\subsection*{Proof of Theorem~\ref{the: sufficient cond}}
\begin{proof}
Let $\tilde p_i=p_i\alpha(\chi)$. Then
\[
\frac{\partial_{\theta_i}\tilde p_i}{\tilde p_i}
=\frac{\partial_{\theta_i}p_i}{p_i}
+\frac{\alpha'(\chi)}{\alpha(\chi)}\,\partial_{\theta_i}\chi
\]
If $\tfrac{\partial_{\theta_i} p_i}{p_i}\ge -\tfrac{\alpha'(\chi)}{\alpha(\chi)}\,\partial_{\theta_i}\chi$ pointwise, then $\partial_{\theta_i}\tilde p_i\ge0$ 
The expected misses are
\[
\bar U(\mathbf{a})
=\mathbb{E}_\omega\!\!\int_{\Psi\times\mathcal{T}}
\lambda(s,t)\prod_{j\in\mathbf{a}}(1-\tilde p_j)\,ds\,dt
\]
whose Gâteaux derivative with respect to $\tilde p_k$ is nonpositive, so $\bar U(\mathbf{a})$ decreases as $\theta_i$ increases. Since $\nu(\mathbf{a})=\exp(-\bar U(\mathbf{a}))$, the void probability increases, proving the claim.
\end{proof}

\subsection*{Proof of Theorem~\ref{the:Uniform}}
\begin{proof}
For $(i,g)$ let $X^{(n)}_{i,g}\in[0,1]$ be Bernoulli trials with mean $\tilde p_{i,g}$ and empirical mean $\widehat p_{i,g}=\tfrac{1}{N}\sum_{n}X^{(n)}_{i,g}$. Hoeffding’s inequality gives
\[
\Pr(|\widehat p_{i,g}-\tilde p_{i,g}|\ge\varepsilon)\le2e^{-2N\varepsilon^2}
\]
A union bound over $mL$ pairs yields
\[
\Pr\!\Big(\max_{i,g}|\widehat p_{i,g}-\tilde p_{i,g}|\ge\varepsilon\Big)\le 2mL\,e^{-2N\varepsilon^2}
\]
Setting the RHS to $\delta$ gives $\varepsilon=\varepsilon_N=\sqrt{\tfrac{1}{2N}\ln\tfrac{2mL}{\delta}}$, proving the result.
\end{proof}

\subsection*{Proof of Lemma~\ref{lemma: error}}
\begin{proof}
Write
\[
\bar U(\mathbf{a})=\sum_g \lambda_g|g|\!\prod_{i\in\mathbf{a}}(1-\tilde p_{i,g}),\quad
\widehat U(\mathbf{a})=\sum_g \lambda_g|g|\!\prod_{i\in\mathbf{a}}(1-\widehat p_{i,g})
\]
Since $f(x_1,\ldots,x_K)=\prod_{j}(1-x_j)$ is $1$-Lipschitz in $\ell_1$,
\[
\Big|\!\prod_{i}(1-\widehat p_{i,g})-\!\prod_{i}(1-\tilde p_{i,g})\Big|
\le K\,\max_i|\widehat p_{i,g}-\tilde p_{i,g}|
\]
Multiplying by $\lambda_g|g|$ and summing over $g$ gives
\[
|\widehat U(\mathbf{a})-\bar U(\mathbf{a})|
\le C_K\,\max_{i,g}|\widehat p_{i,g}-\tilde p_{i,g}|
\le C_K\,\varepsilon_N
\]
with probability $1-\delta$ by Theorem~\ref{the:Uniform}.
\end{proof}

\subsection*{Proof of Theorem~\ref{theo:greedy stability}}
\begin{proof}
For the estimated model $\widehat p$, the greedy bound holds:
\begin{align}
\widehat U(\widehat{\mathbf{a}}_{\mathrm{greedy}})
\le \tfrac{1}{e}\Lambda+ (1-\tfrac{1}{e})\,\widehat U(\widehat{\mathbf{a}}^\star) \label{eq: proof1}  
\end{align}
By Lemma~\ref{lemma: error},
\begin{align}
\widehat U(\widehat{\mathbf{a}}^\star)\le U(\mathbf{a}^\star)+C_K\varepsilon_N \label{eq: proof2}    
\end{align}
and
\begin{align}
U(\widehat{\mathbf{a}}_{\mathrm{greedy}})\le\widehat U(\widehat{\mathbf{a}}_{\mathrm{greedy}})+C_K\varepsilon_N \label{eq: proof3}    
\end{align}
Combining (15)-(17),
\[
U(\widehat{\mathbf{a}}_{\mathrm{greedy}})
\le \tfrac{1}{e}\Lambda + (1-\tfrac{1}{e})U(\mathbf{a}^\star) + (2-\tfrac{1}{e})C_K\varepsilon_N
\]
Setting $C'_K=(2-\tfrac{1}{e})C_K$ and using $\nu(\mathbf{a})=\exp(-U(\mathbf{a}))$ yields the bound in \eqref{eq: lower bound of finite samples}
\end{proof}

\bibliographystyle{IEEEtran}
\bibliography{IEEEabrv,main}

\end{document}